\documentclass{mn2e}
\usepackage{graphicx}

\newif\ifAMStwofonts
\AMStwofontstrue

\def\pg{{PG1211+143}}

\def\xmm{{\it XMM-Newton}}

\def\et{{et al.\ }}

\newcommand{\ls}{\mathrel{\hbox{\rlap{\hbox{\lower4pt\hbox{$\sim$}}}\hbox{$<$}}}}
\newcommand{\gs}{\mathrel{\hbox{\rlap{\hbox{\lower4pt\hbox{$\sim$}}}\hbox{$>$}}}}

\def\arcs{{\hbox{$^{\prime\prime}$}}}

\def\dg{^{\circ}}
\def\H0{{\rm ~km~s^{-1}~Mpc^{-1}}}

\def\le{{L_{\rm Edd}}}
\def\msun{{\rm M_{\odot}}}

\def\mo{{\dot M_{\rm out}}}
\def\me{{\dot M_{\rm Edd}}}

\def\msun{M_{\rm \odot}}

\def\le{{L_{\rm Edd}}}
\def\msun{{\rm M_{\odot}}}

\def\mo{{\dot M_{\rm out}}}
\def\me{{\dot M_{\rm Edd}}}

\def\et{{et al.}}

\def\deg{^\circ}

\title[A high velocity outflow from \pg]
        {Quantifying the fast outflow in the luminous Seyfert galaxy \pg\ }
\author[K.A.Pounds \et]
        {K.A.Pounds,$^{1}$
	J.N.Reeves,$^{2}$\\
$^1$ Department of Physics and Astronomy, University of Leicester,
Leicester, LE1 7RH, UK\\
$^2$ Astrophysics Group, School of Physical and Geographical Sciences, Keele University, Keele, ST5 5BG, UK\\}
\date{Accepted ; Submitted }
\pagerange{\pageref{firstpage}--\pageref{lastpage}}
\pubyear{2008}
\begin{document}
\maketitle
\label{firstpage}

\begin{abstract} 

We report two new \xmm\ observations of \pg\ in December 2007, again finding evidence for the fast outflow of highly
ionised gas first detected in 2001. Stacking the new spectra with those from two earlier \xmm\ observations reveals strong and broad
emission lines of FeXXV and OVIII, indicating the fast outflow to be persistent and to have a large covering factor. This finding confirms a high
mass rate for the ionised outflow in \pg\ and provides the first direct measurement of a wide angle, sub relativistic outflow from an AGN transporting mechanical
energy with the potential to disrupt the growth of the host galaxy. We suggest \pg\ may be typical of an AGN in a rapid super-Eddington growth phase.

\end{abstract}


\section{Introduction}
An early \xmm\ observation of the bright QSO \pg\ in 2001 provided strong evidence of a radial outflow of highly ionised gas 
with the remarkably high velocity of  $\sim$0.13c (Pounds \et\ 2003; Pounds and Page 2006). Unless viewed along the
axis of a highly collimated flow, the high column density required to produce the observed Fe K absorption, combined with the
high velocity, implied a mass outflow rate comparable to the accretion rate and transporting
mechanical energy at a significant fraction of the bolometric luminosity. The broader potential importance of such energetic flows,
which we have suggested might be typical of AGN accreting at the Eddington rate (King and Pounds 2003), is in offering a feedback mechanism
that could link the growth of supermassive black holes in AGN with their host galaxy (Ferrarase and Merritt 2000, Gebhardt \et\ 2000,
Tremaine \et\ 2002, King 2003, Kim \et\ 2008). The greatly improved sensitivity of current X-ray observations in the Fe K energy band
is now yielding more examples of highly ionised outflows with large column densities and high velocities (see Cappi 2006 for a recent review),
but the lack of information on the outflow collimation and covering factors has left large uncertainties in the total mass and energy of these 
flows (Elvis 2006).  

A second \xmm\ observation, in 2004, found \pg\ to be in a brighter flux state, but with both the highly ionised and
the moderately ionised spectral features responsible for the mid-band spectral curvature (and thereby contributing to a strong `soft
excess') significantly weaker.  We now report two further \xmm\ observations in December 2007, finding the integrated X-ray
flux to be higher again than in 2004. However, while the flux increase appears to be dominated by a soft continuum
component, consistent with previous modelling of the broad band X-ray spectrum of \pg\ (Pounds and Reeves 2007; hereafter
P07), the blue-shifted Fe K absorption is again clearly detected. We now combine data from all four observations of \pg\ in an attempt to
characterise and quantify the re-emission from the highly ionised outflow and use this to clarify the flow structure, mass and energy.    

We assume a redshift for \pg\ of $z=0.0809$ (Marziani \et\ 1996).

\section{Observation and data reduction}

The 2007 observations of \pg\ by \xmm\ took place on December 21 and 23, with on-target exposures of $\sim$65 ks and $\sim$50
ks.  In this paper we use data from the EPIC pn  and MOS cameras (Str\"{u}der \et\ 2001,Turner \et\ 2001) and the high resolution 
Reflection Grating Spectrometer (den Herder \et\ 2001). Source
counts from the EPIC cameras were taken from a circular region of $\sim$ 45\arcs\ radius around the centroid position of \pg,
with simultaneous background spectra from a larger region, offset from but close to the target source. X-ray data  were
extracted with the XMM SAS v7.1 software and events selected corresponding to patterns 0-4 (single and double pixel events)
for the pn camera and patterns 0-12 for the MOS cameras. Individual source spectra were integrated over the periods of low
background and binned to a minimum of 20 counts per bin to 
facilitate use of  $\chi^2$
minimalisation in spectral fitting. 

Spectral fitting was based on the Xspec package (Arnaud 1996) and all fits included absorption due to the line-of-sight
Galactic column of $N_{H}=2.85\times10^{20}\rm{cm}^{-2}$ (Murphy \et\ 1996). Errors on individual parameters are quoted at
the 90\% confidence level.

\section{Visual examination of the EPIC data} 
  
Figure 1 displays the new pn spectral data of \pg, together with data from the earlier \xmm\ observations. In comparing such
raw spectra, we note that each represents an  integration of up to a day, with separations of 3 years, 6 years and 2-3 days.
Visual examination of figure 1 suggests, however, that rather clear differences in  such spectral `snapshots' could be
instructive in exploring spectral composition and variability. Thus, the ratio of the 2001 and 2004 EPIC count rate spectra
(black and red in figure 1) led us to conclude (P07) that a decrease in continuum  absorption was the  main cause of the broad
spectral change from 2001 to 2004.  An important constraint on the nature of that variable absorption is the similarity in
the ionising flux during the 2001 and 2004 observations indicated by the near-identical spectra above $\sim$3 keV, suggesting
a change in covering factor rather than ionisation parameter.

\begin{figure}                                                          
\centering                                                              
\includegraphics[width=6cm, angle=270]{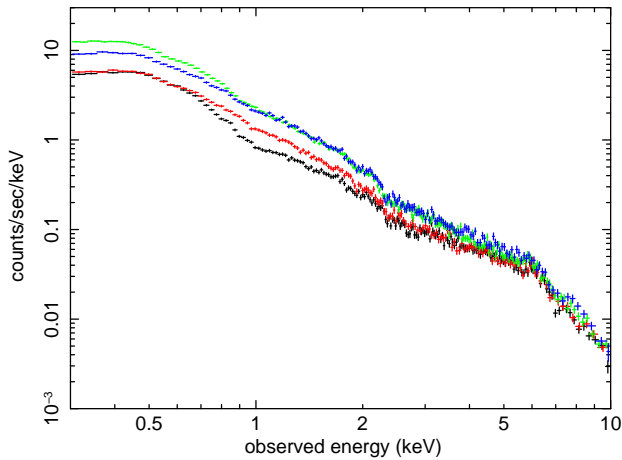}                     
\caption                                                                
{Comparison of the unmodelled pn spectra of \pg\ from the 2001 (black), 2004 (red), 2007a (green) and 2007b (blue) observations.
 As noted previously (Pounds and Reeves 2007) the differences between the 2001 and 2004 spectra can be largely attributed to
 weaker ionised absorption in 2004}
\end{figure}      

In both 2007 observations \pg\ was a factor 2-3 brighter still than in 2004, though again converging at the upper end of the EPIC
energy band.  In modelling the 2001 and 2004 spectra we introduced a second continuum component which
contributed strongly in the soft X-ray band (P07). Direct comparison of the highest (2007a) and lowest  (2001) flux spectra now
allow that model to be tested over a wider flux range. Figure 2 shows the outcome, with the count rate difference spectrum
being well fitted by a power law of $\Gamma$$\sim$3, unaffected by continuum absorption. 

We leave further discussion of this soft continuum component to another time while noting that - if a common property of luminous AGN -
a soft, variable continuum substantially reduces the problem of the `soft excess' widely debated over many years 
(e.g. Sobolewska and Done 2007 and references therein).

\begin{figure}                                                                              
\centering                                                              
\includegraphics[width=6cm, angle=270]{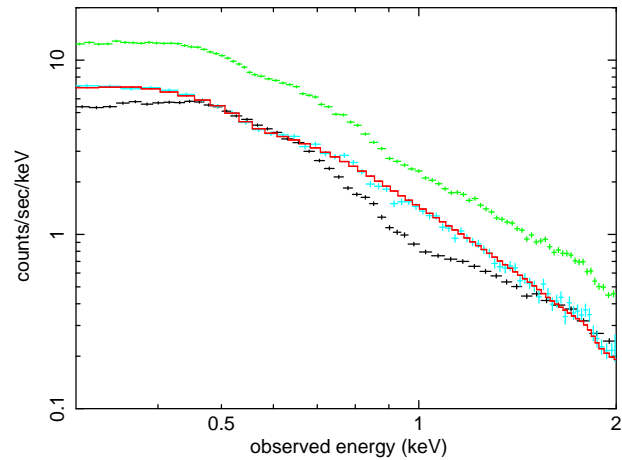} 
\caption                                                                
{pn data from the 2001 (black) and 2007a (green) observations and the difference spectrum (light blue) obtained by a direct
subtraction of the data sets. Modelling the difference spectrum finds a simple power law of $\Gamma$$\sim$3 (red) to provide
a good fit over the whole 0.3-10 keV EPIC band}
\end{figure}

\subsection{Further evidence for a high velocity outflow}

The most important outcome of the 2001 observation of \pg\ was the detection of blue-shifted absorption lines of highly ionised Fe, 
S and Mg, initially
interpreted (Pounds \et\ 2003) as arising in highly ionised gas outflowing at $\sim$0.1c. A more complete analysis of pn
and MOS spectra, based on the detection of 7 absorption lines of Ne, Mg, Si, S and Fe,  removed an ambiguity on line
identification,  confirming a lower ionisation parameter and correspondingly higher velocity of $\sim$0.13c (Pounds and Page 2006). Emission lines were less well
defined, and the model of P07 offered a possible explanation in terms of strong velocity broadening. The new observations 
in 2007 now provide the opportunity to test that proposal and further explore the properties of the ionised outflow in \pg.

Figure 3 shows the Fe K emission/absorption profiles from the 2001 pn data and the sum of the two 2007 observations. Each
profile is the ratio of observed counts to an underlying continuum modelled by an absorbed power law over the 3-10 keV
band. The data have been binned to minima of 50 and 100 counts, respectively, to improve the statistics in the highest
energy channels.  While the overall profiles in figure 3 are similar, with significant
emission to the low energy side of the $\sim$7 keV absorption line in both data sets, the statistical quality of the individual 
profiles is clearly limited.   

Obtaining a better-defined Fe K profile is important as the strength and width of an ionised emission component
potentially carries crucial information on the covering factor and collimation of the outflow.  That, in turn, is the key to
determining the mass rate and mechanical energy in an outflow whose velocity is known. As we are interested in
the time-averaged emission to assess the  covering factor of the ionised outflow, and noting the essentially constant
underlying continuum in the Fe K energy band (figure 1), it should be appropriate to utilise the integrated spectrum from all
four \xmm\ observations. 
In stacking the data from the different time periods we used a weighted mean response function and
adjusted the background scaling factor to account for the actual image areas used to collect individual data sets.

The Fe K profile from the stacked pn data is illustrated in figure 4 and resembles the PCygni profile characteristic of an outflow, 
with both 
emission and blue-shifted absorption components now better defined. While integrating the data from four observations, over 6 years, is likely to blur narrow or transient features,
the improved statistics should offer a better measure of the average outflow properties.

We analyse the PCygni profile in Section 5. 
Before that, the spectral model developed in P07 is re-fitted to the stacked data over the whole EPIC band in order
to characterise the broad band spectrum and possibly identify the spectral features that could contribute to the Fe K emission/absorption 
profile.

\begin{figure}                                                                              
\centering                                                              
\includegraphics[width=6cm, angle=270]{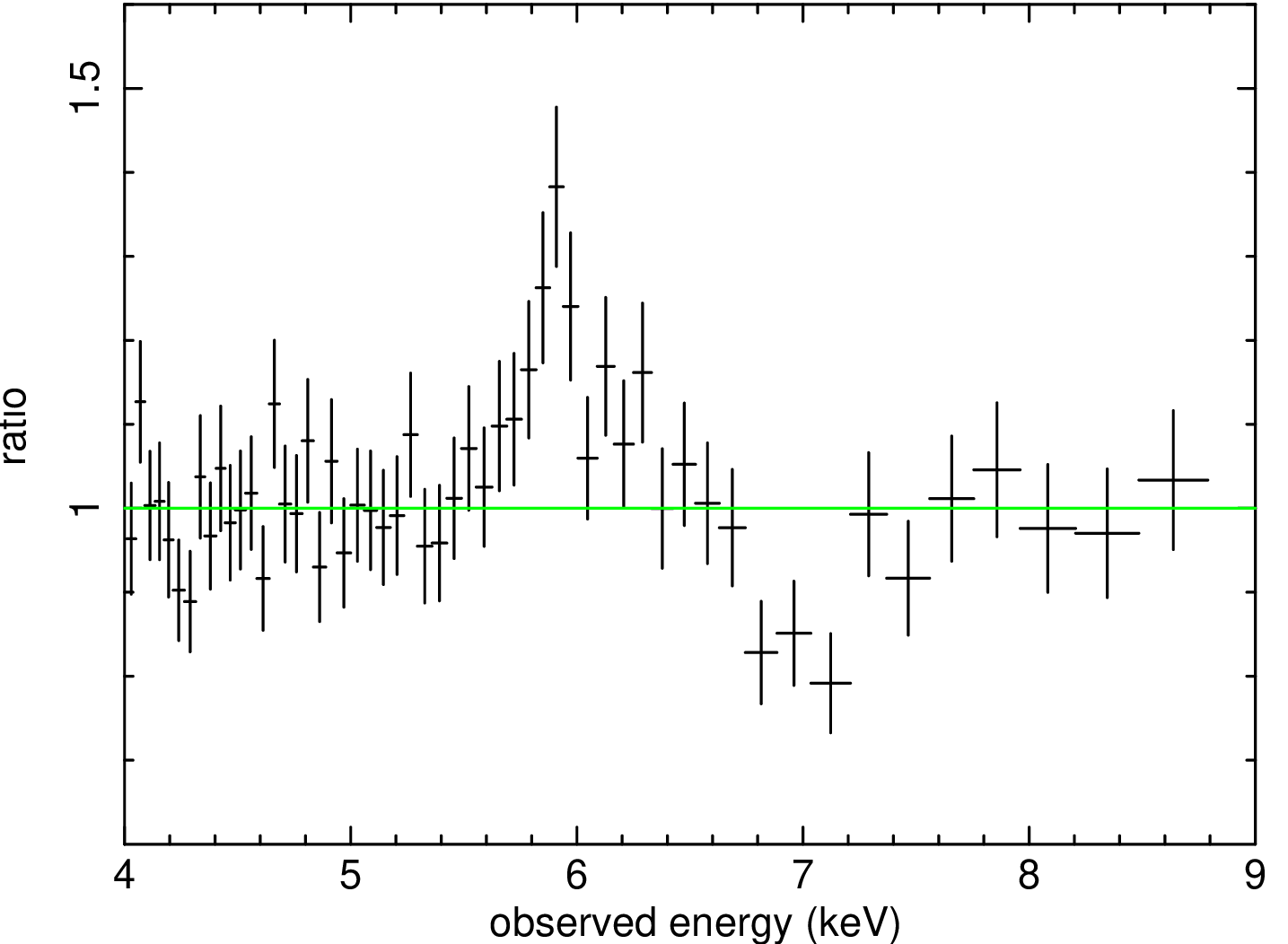} 
\centering                                                              
\includegraphics[width=6cm, angle=270]{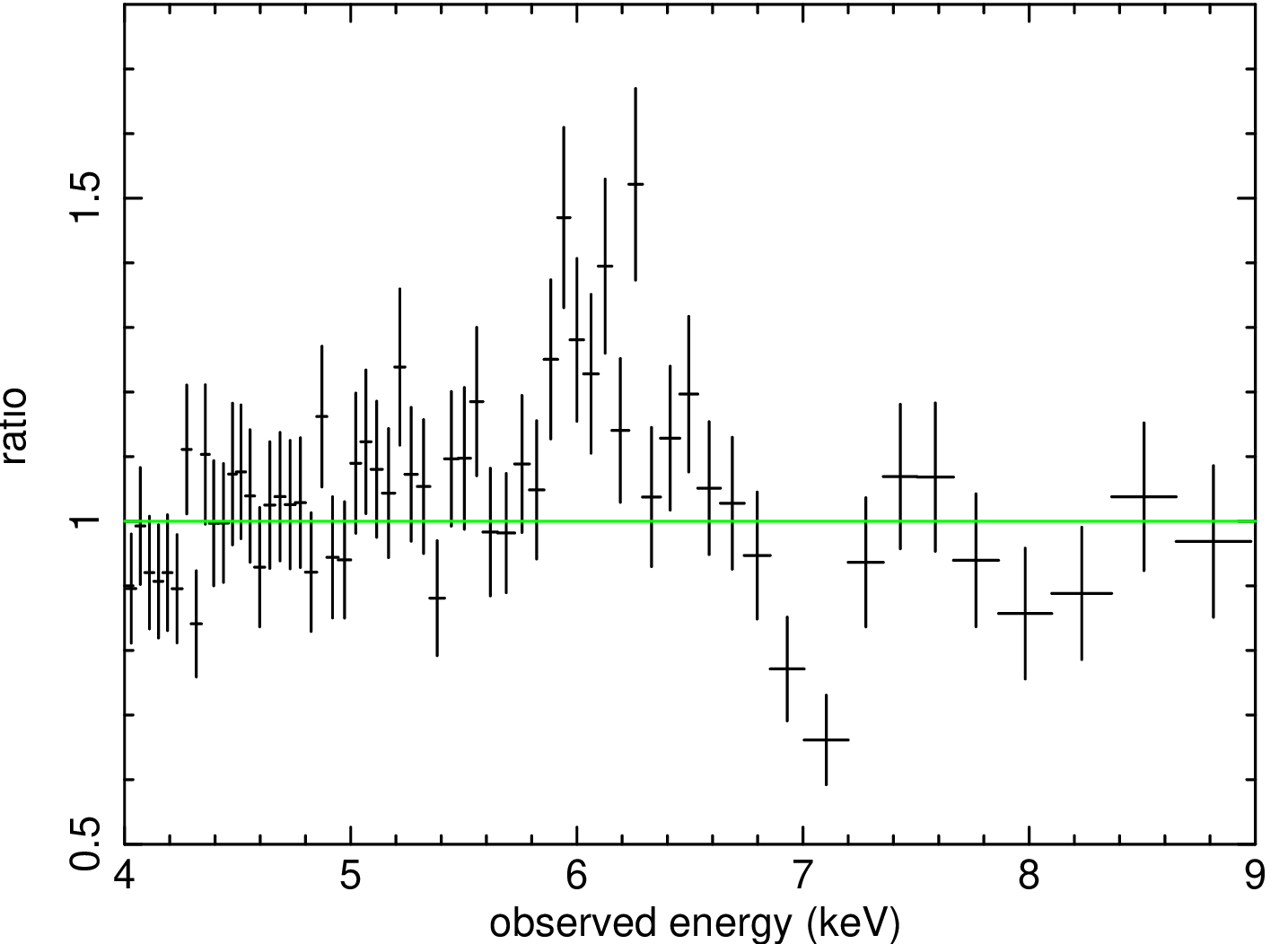} 
\caption                                                                
{Fe K profiles expressed as a ratio of spectral data to a best-fit continuum from the 2007 (top) and the 2001 (lower) pn observations of \pg. While a narrow
absorption line is seen near $\sim$7 keV in in both 
profiles the emission to lower energies is not well defined}
\end{figure}   
  
\begin{figure}                                                                              
\centering                                                              
\includegraphics[width=6.3cm, angle=270]{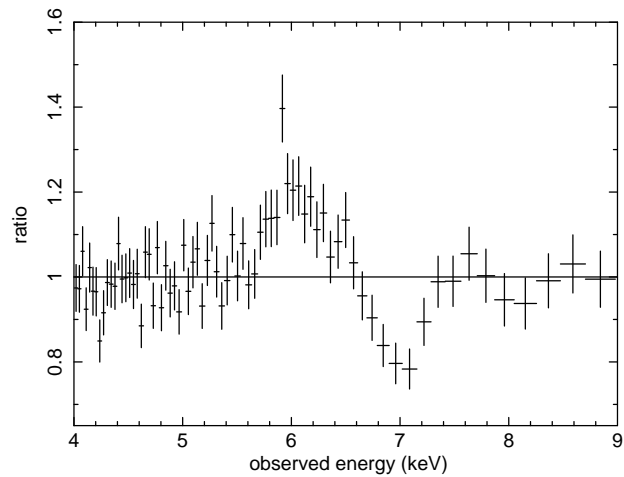} 
\caption                                                                
{Fe K profile from the stacked pn observations of \pg. The PCygni-like profile shows the emission and blue-shifted absorption characteristic
of an outflow are more clearly defined}
\end{figure}

\section{Modelling the overall X-ray spectrum}

The spectral model of P07 provided a physically consistent description of the broad band X-ray spectra of \pg\  observed by
\xmm\ in 2001 and 2004. It has the form in Xspec of  wa(po1*abs1 + g1*em1 + g2*em2 + po2)abs2, with the `primary' power law (po1) modelling the
energetically dominant continuum and a steeper continuum component (po2) found to dominate the soft X-ray variability.  Photoionised absorption
(abs1,2) and emission (em1,2) from highly and moderately ionised gas are modelled with the publicly available XSTAR grid 25, based on the revised
treatment of Fe K (Kallman \et 2004). Grid 25 has an assumed turbulence velocity of 
200 km s$^{-1}$ \footnote{A higher turbulent velocity would result in lower column densities but would not significantly alter the main results of this
analysis } and element abundances fixed at solar values (Grevesse and Sauval 1998).  Free
parameters in fitting the absorption and emission spectra are the ionisation state, column density and apparent redshift, the latter being an indicator of the
velocity of each photoionised gas component. Line broadening is allowed for in the model fit by convolving a separate Gaussian (g1,g2, modelled 
by gsmooth in Xspec) with each emission line
spectrum. 

Applying this spectral model to the stacked pn and MOS data, separately, over the whole EPIC band yielded very similar fits, with the principal parameters 
listed in Table 1. The unfolded spectrum for the pn data is illustrated in figure 5. 

The primary power law component (po1) has a photon index of $\Gamma$$\sim$2-2.1 and is strongly absorbed by moderately ionised gas of column density
$N_{H}$$\sim$$8\times 10^{22}$ cm$^{-2}$  and ionisation parameter log$\xi$$\sim$2.0 erg cm s$^{-1}$. The secondary power law (po2) 
has a much steeper slope ($\Gamma$$\sim$2.8-2.9) and is affected only by the more highly ionised absorber.

\begin{table*}
\centering
\caption{Parameters of the model fit to the stacked pn and MOS data from the 2001, 2004 and 2007 \xmm\ observation of \pg. Power law indices
$\Gamma$$_{1}$ and  $\Gamma$$_{2}$ refer to the primary and secondary continuum components, respectively. High and moderate
ionisation absorbers affect the primary continuum, with equivalent hydrogen column densities in units of $10^{22}$ cm$^{-2}$
and ionisation parameters in erg cm s$^{-1}$. The soft secondary continuum is affected only by the high ionisation absorber.
Apparent redshifts are converted to ionised gas outflow velocities in the text}
\begin{tabular}{@{}lcccccccccc@{}}
\hline
&$\Gamma$$_{1}$ & $\Gamma$$_{2}$ & $N_{H}$ & log$\xi$ & redshift & $N_{H}$ & log$\xi$ & redshift & $\chi^{2}$/dof \\
\hline
pn & 2.09$\pm$0.09 & 2.92$\pm$0.02 &  7.8 & 2.04$\pm$0.02 & 0.022$\pm$0.001   & 82 & 3.37$\pm$0.01 & -0.026$\pm$0.004 & 548/478 \\
\hline
MOS & 2.03$\pm$0.10 & 2.78$\pm$0.04 &  7.7 & 2.03$\pm$0.03 & 0.010$\pm$0.001   & 90 & 3.36$\pm$0.01 & -0.024$\pm$0.01 & 398/354 \\
\hline
\end{tabular}
\end{table*}

The highly ionised absorber has a well constrained ionisation parameter in both pn and MOS fits, with log$\xi$ $\sim$3.4 and a large column density  
N$_{H}$$\sim$$8\times10^{23}$cm$^{-2}$. The 
`apparent redshift' is $\sim$ -0.026, implying the absorbing gas is strongly blue-shifted in the rest frame of \pg. As the redshift measure is dependent on 
the primary line identification (and hence ionisation parameter)
in the modelling, these important parameters are not independent. To explore this dependence we obtained confidence contours for the ionisation parameter  and
redshift of the highly ionised absorber component with all other model parameters free. Figure 6 shows the outcome of this check, confirming an ionisation parameter 
of log$\xi$
$\sim$3.37 and apparent redshift of -0.025$\pm$0.005.  Converting apparent redshift 
to the rest frame of \pg\ gives an outflow velocity for the highly ionised gas of 
$\sim$0.099$\pm$0.005c. The precision of this velocity measure, which derives from fitting to the strong Fe XXV line and an array of co-moving
absorption lines best seen in the
2001 data (Pounds and Page 2006), is probably misleading given the assumption of a unique ionisation parameter for the fast outflow.     

The model fit 
for the moderate ionisation absorber yielded an apparent redshift of 0.02$\pm$0.01 and a significantly lower outflow velocity of
$\sim$0.06$\pm$0.01c.  
\footnote{For ionisation parameters in the range log$\xi$ $\sim$1.8-2.4 the XSTAR models predict inner shell (1s-2p) absorption at $\sim$6.5 keV from FeXVIII-XXIII (Behar
and Netzer 2002). For an outflow velocity 
of $\sim$0.06c this feature would be observed at $\sim$6.3 keV in \pg. Intriguingly, figure 7 does show a small dip at 6.35$\pm$0.05 keV, although  
this is
not statistically significant in the present data}

The photoionised emission spectra were found to have ionisation parameters of log$\xi$$\sim$3.2 and log$\xi$$\sim$1.2, both with apparent
redshifts of 0.08$\pm$0.01, indicating a low mean velocity  relative to the AGN. Importantly, the coincidence of absorption and emission ionisation
parameters for the fast outflow implies these are from the same flow component. In contrast, 
significant re-emission from the moderately ionised gas was not required in the spectral fit, suggesting it has a small emission measure.  

Importantly, strong broadening of the high ionisation emission lines is implied by the parameter g1, for which a value $\sigma$$\sim$340$\pm$180 eV at 6 keV 
(with constant $\Delta E /E$ ) significantly improved
the quality of the fit (reducing $\chi^{2}$ by 17 for 1 fewer d.o.f.), while allowing the line fluxes to increase by a factor $\sim$5. In contrast, the
Gaussian broadening parameter  for the low ionisation gas, $\sigma$$\sim$60 eV, was found to make very little difference to the overall spectral fit. 

Visual examination of figure 5 suggests that only the
highly ionised gas (light blue in the figure) is likely to contribute significant line emission in the Fe K band, with a strong  resonance  line
of He-like FeXXV (rest energy 6.7 keV). Blue-shifted FeXXV is also identified with the strongest absorption line arising from the highly ionised gas,
while the lower ionisation column results in a significant Fe K absorption edge and curvature in the primary (po1) continuum. 

It is particularly interesting that the spectral fit requires
the highly ionised emission line spectrum to be strongly broadened. Deconvolution of the Fe K profile in the stacked data 
offers the promise of a direct measure of this broadening in the case of a dominant  Fe XXV emission line, thereby providing a key indicator of the geometry of
the outflow. 

\begin{figure}                                                                              
\centering                                                              
\includegraphics[width=6.3cm, angle=270]{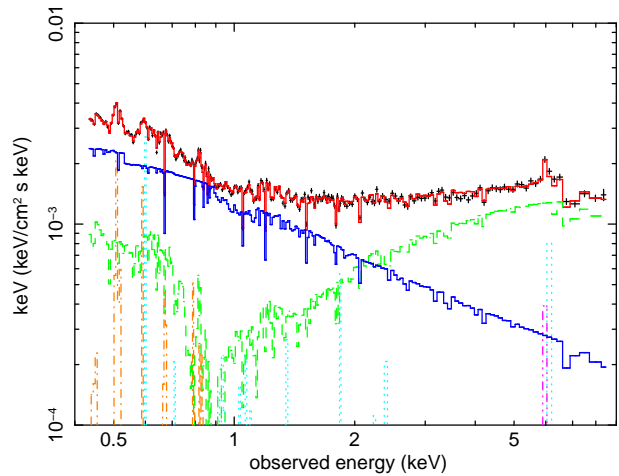} 
\caption                                                                
{Unfolded broad band model fitted to the stacked pn spectral data. The continuum components are shown in green and dark blue, and the ionised gas
emission spectra in light blue and brown.  A narrow emission line (pink) has been added to represent the feature found in the Gaussian fitting
described in the next section and attributed to Fe K-$\alpha$ fluorescence from cold matter }
\end{figure}

\begin{figure}                                                                              
\centering                                                              
\includegraphics[width=6cm, angle=270]{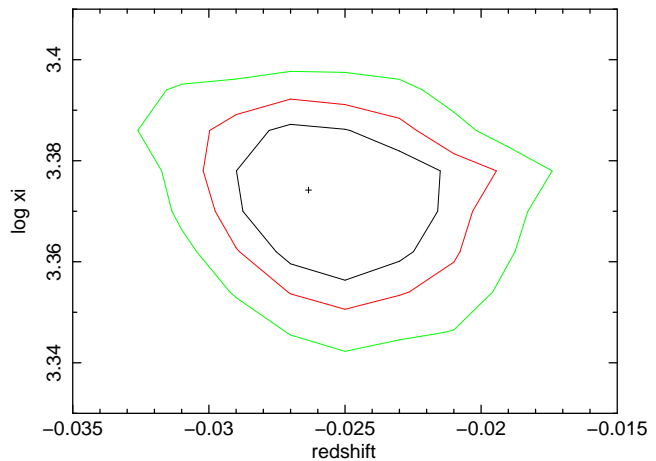} 
\caption                                                                
{68, 90 and 99 percent confidence contours for the ionisation parameter and apparent redshift of the highly ionised component of the outflow, assessed
 with all model parameters free in the pn data fit}
\end{figure}

\section{Resolving the PCygni profile of Fe K in the EPIC data} 

Figure 4 shows the profile of Fe K emission and absorption for the stacked pn data, plotted as a ratio to the best-fit continuum. Guided by the 
broad band model fitted in the previous section, the continuum is modelled over the 3-10 keV band with the form in Xspec wa(zwa*po1 + po2), where zwa allows for
absorption of the primary power law at the redshift of \pg. A best-fit continuum was found for power law indices of $\Gamma$=2.2 and 2.9, with
a cold absorbing column of $10^{23}$ cm$^{-2}$. Tests showed these parameters are important in determining the details of the PCygni profile deconvolution,
mainly in the deduced line strengths, but do not change
the need for two dominant components, one in absorption at $\sim$7 keV and a broader emission component at a lower energy.   

We initially analysed the PCygni line profile by adding a sequence 
of positive and negative Gaussians to produce a visual and quantitative fit. The main profile was found to be well fitted by the addition of just 2 lines,
both broader than the resolution of the pn camera. Figure 7 shows this fit, with
emission and absorption component energies (adjusted to  the redshift of \pg) of 6.51$\pm$0.04
keV and 7.65$\pm$0.05 keV, and 1 sigma line widths (which here include $\sim$65 eV for the pn resolution) of 210$\pm$50 eV and 170$\pm$50 eV, respectively. The statistical 
improvement by adding these 2 Gaussians
is listed, together with the individual component parameters, in Table 2.

Although already a good fit, the emission component in figure 7 has an obvious high data point which lies close to 5.92 keV, which corresponds 
(at the redshift of
\pg) to the 6.4 keV Fe K$\alpha$ line, often seen in the spectra of AGN and attributed to fluorescence from low velocity (distant?) cold matter. 
Adding a third Gaussian to the profile fit, 
with width constrained to the pn energy resolution, gave a further small but significant improvement to
the fit, while increasing the energy of the broad emission
component to 6.61$\pm$0.08 keV and its width to 260$\pm$60 eV. The absorption line parameters were essentially unchanged. Figure 8 reproduces this
3-Gaussian fit with parameters summarised in Table 2. 

\begin{table*} 
\centering
\caption{Parameters of Gaussian absorption and emission lines sequentially fitted to the ratio of stacked pn data to continuum over the 3-9 keV
band.
Line energies (adjusted to the AGN redshift) and 1$\sigma$ line widths are in keV and equivalent widths in eV. The final column 
gives the improvement in $\chi^{2}$ for each successive fit}
\begin{tabular}{@{}lcccccccccc@{}}
\hline
fit & energy & width & EW & energy & width & EW & energy & width & EW  & $\Delta\chi^{2}$/dof \\
\hline
1 & 7.60$\pm$0.05 & 0.19$\pm$0.06 & -120$\pm$30 &  &  &  &  &  & &   49/3 \\
\hline
2 & 7.65$\pm$0.05 & 0.17$\pm$0.05 & -110$\pm$40 & 6.51$\pm$0.04 & 0.21$\pm$0.05 & 150$\pm$45 &  &  & &   47/3 \\
\hline
3 & 7.65$\pm$0.05 & 0.16$\pm$0.06 & -110$\pm$45 & 6.61$\pm$0.08 & 0.26$\pm$0.06 & 135$\pm$45 & 6.4 & 65 & 25$\pm$10    & 5/1 \\
\hline
\end{tabular}
\end{table*}

The broad emission component is now consistent with the resonance 1s-2p transition in He-like FeXXV, the same transition
as identified in the absorption spectrum in the 2001 data (Pounds and Page 2006). Allowing for the pn camera
resolution, an intrinsic emission line width of 250$\pm$60 eV, or 28000$\pm$7000 km s$^{-1}$ (FWHM) when interpreted solely in terms of 
velocity broadening,
corresponds to a wide angle outflow of semi-angle $\sim$50 $\deg$. With the reasonable assumption that the FeXXV emission and absorption lines
come from the same ionised outflow, the comparable equivalent widths of the two
components is further strong evidence for the outflow having a large covering factor.  

The significantly narrower absorption line in the stacked data constrains the effects of velocity variations and (unknown) turbulence, while also ruling out a
significant change in the ionisation state of the highly ionised gas, for example from dominantly FeXXV to FeXXVI. We conclude that the greater broadening of the emission line with respect to the 
absorption line in the stacked data is most likely due to
the spread of projected velocities in a wide angle flow.

To check the robustness of the broad ionised emission line in the stacked data, an alternative fit was tried with 3 narrow lines
corresponding to neutral Fe K-$\alpha$ (rest energy 6.4 keV), and the resonance emission lines of FeXXV (6.7 keV) and FeXXVI
(6.97 keV). The fit was noticeably worse ($\Delta\chi^{2}$= 38) than that with the narrow Fe K-$\alpha$ plus broad FeXXV 
emission line (figure 9). 

\begin{figure}                                                                              
\centering                                                              
\includegraphics[width=6cm, angle=270]{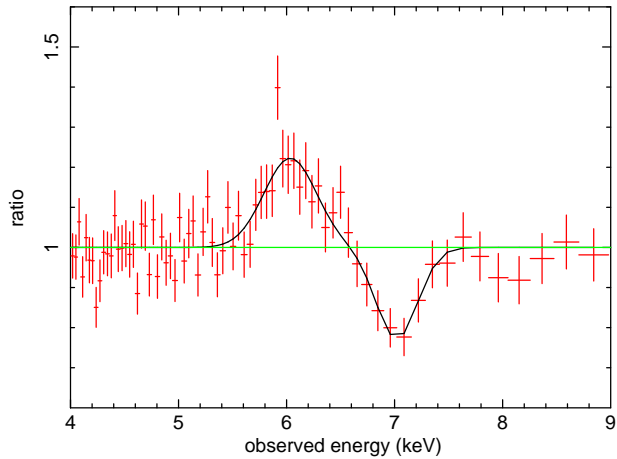} 
\caption                                                                
{Double Gaussian line fit to the stacked pn data Fe K profile showing a classical P Cygni profile}
\end{figure}

\begin{figure}                                                                              
\centering                                                              
\includegraphics[width=6cm, angle=270]{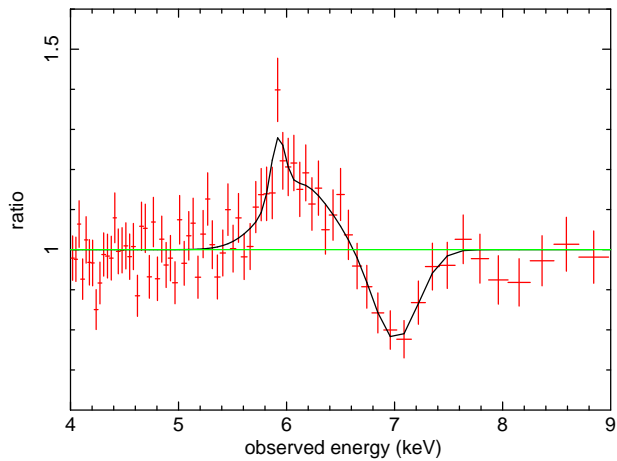} 
\caption                                                                
{Figure 7 with an additional narrow emission line at the rest energy of Fe K-$\alpha$}
\end{figure}

\begin{figure}                                                                              
\centering                                                              
\includegraphics[width=6cm, angle=270]{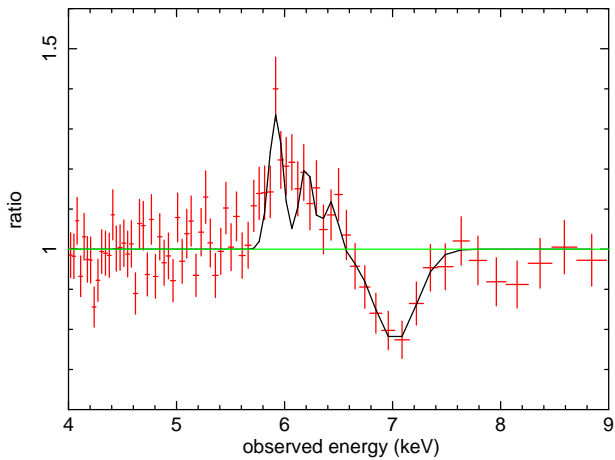} 
\caption                                                                
{An alternative Gaussian fit to the Fe K profile with narrow lines corresponding to neutral Fe K-$\alpha$ (rest energy 6.4 keV) and resonance
emission lines of FeXXV (6.7 keV) and FeXXVI (6.97 keV)}
\end{figure}

Finally, to confirm the broad emission line is not an artefact of the stacking procedure the 4 pn data sets (figure 12 in the Appendix) were fitted 
individually. 
Although less well determined, a broad Fe K emission line is preferred in each individual data set, with widths and line energies consistent, within the errors,
with those derived for the 
stacked data (Appendix, table 4). 

In summary, analysis of the Fe K line profile provides evidence for a velocity-broadened emission line at a rest energy consistent with the principal
resonance transition of FeXXV. As FeXXV is also identified as the ionisation state  in the highly ionised absorber, it is reasonable to interpret the
emission component in the  PCygni profile with the same outflow. We note that this conclusion, including the emission line broadening, is also consistent
with the description of the broad band X-ray  spectrum of \pg\  in Section 4.

\section{A check for consistency in the RGS spectrum}

The unfolded spectrum in figure 5 suggests that discrete spectral features might also be visible in the soft
X-ray band. If so, the higher resolution of the \xmm\ RGS
could provide a further constraint on  
the broadening of emission lines from the highly ionised
outflow, as H-like ions of O and Ne will co-exist with He-Like FeXXV.

To check this possibility the spectra from all four RGS observations were stacked in the same way as for the EPIC data, and the ratio derived of
the background-subtracted data to a best-fit power law (we found $\Gamma$$\sim$2.9, in agreement with the po2 component in Table 1) over the 14-28 Angstrom band. 
The data were then re-grouped in 9-channel bins to 
reduce
random noise, giving an effective spectral resolution of $\sim$0.1 Angstrom. 

Figure 10 shows the resulting RGS data:continuum spectral ratio and marks the rest wavelengths of the
principal emission lines of OVII and OVIII, together with the edge energies of the recombination continua of OVII and NVI. 
As with the Fe K line profile, Gaussians were fitted to the 3 most obvious emission lines, with wavelength, width and amplitude 
as free parameters.

Of
particular interest is a well-defined - and strong - OVIII Lyman-$\alpha$ line, as predicted in figure 5.  The Gaussian fitting finds a wavelength of 
20.48$\pm$0.08 Angstrom  (18.95$\pm$0.08 Angstrom at the redshift of \pg), with a line width of $\sigma$=0.42$\pm$0.08 Angstrom.
Interpreted in terms of velocity broadening, the line width corresponds to $\sim$ 14000$\pm$5500 km s$^{-1}$ (FWHM).
While significantly lower than the value found from fitting the FeXXV emission line in the EPIC data, 
the higher intrinsic resolution of
the RGS does hint at a narrower core and broader wings than  modelled by the simple Gaussian line. Reference to figure 5 suggests that the narrow core might be due 
to a 
contribution from the lower
ionisation, lower velocity gas. 

The OVII (1s-2p) resonance line is also clearly detected in the stacked RGS data, at 23.22$\pm$0.06 Angstrom (21.48$\pm$0.05 Angstrom), but is significantly
less wide than the higher ionisation OVIII line, with $\sigma$=0.16$\pm$0.05 Angstrom. Again, this result is consistent with the narrower 
Gaussian smoothing factor
required for emission from the less ionised gas in the model of figure 5.  Also seen in figure 10 is an unresolved line very
close to the rest wavelength of the forbidden line in the OVII 1s-2p triplet. We surmise that this emission arises from a low
density component of the outflow, presumably at a substantially larger radial distance than the velocity-broadened lines.

Two other spectral features in figure 10 are identified with radiative recombination continua (RRC) of OVII and NVI. Fitting a more
appropriate delta function in Xspec yields an electron temperature of $\sim$0.03$\pm$0.01 keV for the stronger OVII
RRC. 

Table 3 lists the parameters of all 5 spectral features identified in the stacked RGS data, giving an overall improvement to the power law fit 
of $\Delta\chi^{2}$ = 102 for 15
additional degrees of freedom. Figure 11 shows the contribution of the 5 components to the overall spectrum.   

In summary, the stacked RGS data confirm significant photoionised emission in the soft X-ray spectrum of \pg\ as indicated in the spectral 
model of figure 5, and provide further support (particularly from the strength and broadening of the OVIII Lyman-$\alpha$ line) for the
large covering factor of the 
highly ionised
outflow implied by the strong and broad FeXXV resonance emission in the EPIC data. 

\begin{table*} 
\centering
\caption{Parameters of the principal emission features in the stacked RGS data.
The emission line and RRC edge wavelengths (adjusted to the AGN redshift) are in Angstrom with 1$\sigma$ line widths in km s$^{-1}$ (FWHM)}
\begin{tabular}{@{}lcccccc@{}}
\hline
feature & obs.wavelength & lab wavelength & 1$\sigma$ line width & RRC kT (eV) & EW (eV) & $\Delta\chi^{2}$/ dof \\
\hline
OVII RRC & 17.2$\pm$0.1 & 16.80& & 30$\pm$9  & 12.6$\pm$3.5 & 38/3 \\
\hline
OVIII Lyman-$\alpha$ & 18.95$\pm$0.1 & 18.97 & 14000$\pm$3000 & & 11.8$\pm$3.8 & 36/3 \\
\hline
OVII 1s-2p (r) & 21.5$\pm$0.1 & 21.60 & 5000$\pm$1600 & & 2.4$\pm$1.1 & 11/3 \\
\hline
OVII 1s-2p (f) & 22.05$\pm$0.05 & 22.1 & 1500$\pm$900 & & 1.7$\pm$1.0 & 10/3\\
\hline
NVI RRC & 22.45$\pm$0.15 & 22.46 & &15$\pm$10 & 3.7$\pm$2.0 & 7/3 \\

\hline
\end{tabular}
\end{table*}

\begin{figure}                                                                              
\centering                                                              
\includegraphics[width=6.2cm, angle=270]{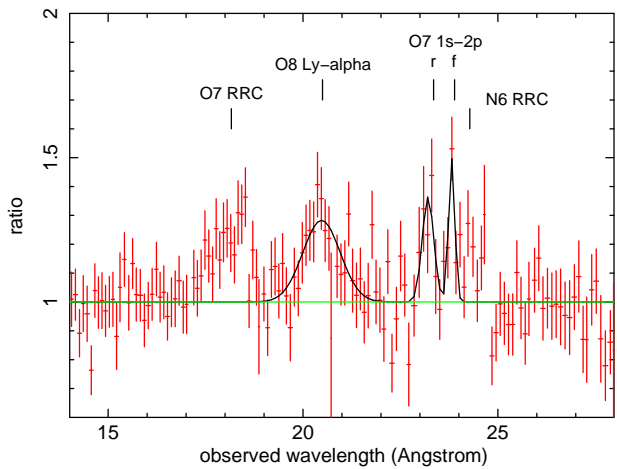} 
\caption                                                                
{Gaussian fits to the stacked RGS data show a broad emission line of OVIII Lyman-$\alpha$ and a less broad OVII 1s-2p resonance line. 
Also shown is a narrow emission line identified with the forbidden transition in OVII and RRC of
OVII and NVI. The rest 
wavelengths of the 
identified features are indicated}
\end{figure}   

\begin{figure}                                                                              
\centering                                                              
\includegraphics[width=6cm, angle=270]{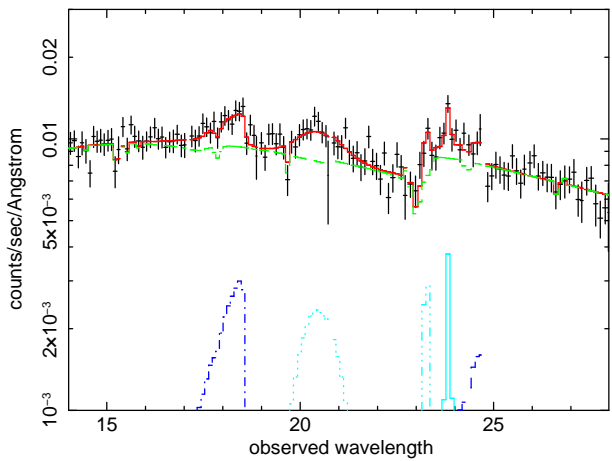} 
\caption 
{The stacked RGS data compared with a power law of $\Gamma$$\sim$2.9 (green) together with the spectral features (blue) 
detailed in Table 3}                                                               
\end{figure}

\section {Discussion}

\subsection {A persistent high velocity outflow}

A significant, but variable strength absorption line has now been observed at $\sim$7 keV in 3 of the 4 \xmm\ observations of \pg, and also in an 
observation from Suzaku (Reeves \et\ 2008). The most recent \xmm\ observations, reported here, show the sub-relativistic, highly
ionised  outflow, originally discovered in the 2001 \xmm\ observation, to be persisting. 

Identifying the $\sim$7 keV absorption line with FeXXV, as in the initial 2001
spectrum (Pounds and Page 2006), is consistent with the ionisation parameter derived from modelling the broad band stacked data.
That model, outlined in Section 4, finds a mean velocity for the highly ionised outflow of $\sim$0.10c, from both pn and MOS fits. Gaussian fitting to the Fe K profile in
Section 5 finds a slightly higher absorption line energy, and implied outflow velocity of $\sim$0.12c. In the following discussion we assume a mean outflow 
velocity for the stacked data of 0.11c. 

Evidence for actual variability in the outflow velocity is offered by the absorption  component being
broader in the stacked data than in the initial observation of 2001.  On that interpretation the measured
width of $\sigma$=160$\pm$60 eV, or 145$\pm$60 eV allowing for the pn resolution, corresponds to a velocity spread of $\pm$0.02c. 

The persistence of the blue-shifted absorption over a 6 year period suggests, unless the flow is highly collimated or
recombination times very short, that  
we should expect to find evidence for an extended region of re-emission
integrated over the outflow. Detecting broad and strong emission lines of FeXXV and OVIII
from the same
highly ionised gas now provides confirmation of that expectation and provides the key information lacking in previous observations of high velocity outflows, namely
the covering factor. In turn, the covering factor allows the actual mass and energy in the flow to be quantified. 

\subsection{An energetically important outflow}

We now combine the fitted parameters for the stacked data from
Sections 4, 5 and 6 to re-assess the structure and hence the wider significance of the fast outflow in \pg. 

We have 3 ways of assessing the flow geometry or
covering factor of the highly ionised gas. The most direct is to compare the strength of the FeXXV absorption and emission components in the PCygni profile.  
The similar equivalent widths of the main absorption and emission components in the FeK profile deconvolution in Section 5 indicates a covering factor 
CF$\sim$1 if the emission is simply from resonance line absorption and re-emission (scattering). While this estimate is
subject to some uncertainty due to the different timescales of the absorption and emission measures, and to the density-dependance of an additional recombination emission 
component, a highly collimated flow is strongly ruled out, with a conservative CF$\sim$0.5.

A second estimate of the CF can be obtained, as in P07, by comparing the total continuum energy absorbed and re-emitted by the highly ionised gas components 
in modelling the stacked
broad band EPIC data. For the pn data we find the luminosity of the highly ionised line
emission spectrum to be 4.3$\times$$10^{42}$~erg s$^{-1}$ while the energy absorbed from the continuum by the highly ionised gas 
is $\sim$1.3$\times$$10^{43}$~erg s$^{-1}$, yielding a CF$\sim$0.33. The corresponding
quantities in the MOS spectral fit are 9.8$\times$$10^{42}$~erg s$^{-1}$ and 1.7$\times$$10^{43}$~erg s$^{-1}$ giving a CF$\sim$0.57.
While these estimates are derived from a rather complex model, we believe they are reliable to a factor 2. 

The third signature of flow geometry is from the observed emission line widths. In that respect it is significant that an acceptable fit 
to the broad band EPIC data required the highly ionised emission lines to be broad. Only by including the Gaussian broadening parameter in the
spectral fitting was the emission line spectrum allowed to increase in strength to match the spectral structure in the data. For the pn data in
Section 4 we
found a Gaussian width of $\sigma$$\sim$340$\pm$180 eV (at 6 keV), consistent with the value $\sigma$$\sim$250$\pm$60 eV derived from the line
profile fitting in
Section 5. (We recall the emission line component in the Fe K profile in a 2005 Suzaku observation of \pg\ appears very similar to that in the stacked \xmm\
data, with a mean energy of $\sim$6.5 keV and width of $\sigma$$\sim$250eV (Reeves \et\ 2008)). 
Assuming a Gaussian width of $\sigma$$\sim$250$\pm$60 eV and the assumption of velocity broadening in a radial flow then corresponds to a flow cone of half
angle $\sim$50 $\dg$ and CF $\sim$0.3.

Importantly, all the above indicators show the highly ionised outflow is {\it not} highly collimated, and therefore involves a significant mass loss.
The mass outflow rate in a uniform radial outflow of velocity v is given by $\mo$ = 4$\pi$bnr$^{2}$m$_{p}$v, where b is the covering fraction,
n is the gas density at
a radial distance r, and m$_{p}$ is the proton mass. We obtain nr$^{2}$ (= L$_{ion}$/ $\xi$) from   
the broad-band model of Section 4 which has a relevant ionising X-ray luminosity L($\ge$7 keV) of 2$\times$$10^{43}$~erg s$^{-1}$
and an ionisation parameter $\xi$(=L/nr$^{2}$)$\sim$2350, yielding nr$^{2}$ $\sim$$8.5\times10^{39}$ cm$^{-1}$. 
For a mean velocity  of 0.11c, the  mass loss rate is then $\mo$ = $\sim$$5.8b\times10^{26}$ gm s$^{-1}$
($ \sim 8.7b\msun$~yr$^{-1}$). 

Assuming from the above estimates a value of b=0.4, we find an average outflow mass rate over the four \xmm\ observations of $\mo$
$ \sim 3.4\msun$~yr$^{-1}$. This compares with $\me$ = 1.6$\msun$~yr$^{-1}$ for a non-rotating supermassive black hole of mass
$\sim$$4\times 10{^7}$$\msun$ (Kaspi \et\ 2000) accreting at an efficiency of 10\%.  The average mechanical energy in the fast outflow is then
of order $\sim$$1.3\times 10^{45}$~erg s$^{-1}$. 

\subsection {A continuum driven outflow from a super-Eddington AGN}

King and Pounds (2003), hereafter K03, described a simple model in which black holes accreting at or above the Eddington rate can drive winds which are, at small
radii, Compton thick in the continuum. It seems likely that this mechanism is responsible for the highly ionised outflow in \pg, as the line driving
favoured
to explain outflows in BAL is ruled out for such highly ionised gas. 

The assumption in K03 is that the matter decouples from the photons at the boundary of the Compton thick flow (the photosphere) and is launched with
an outflow velocity equal to the escape velocity at that radius $R_{launch}$. Here we have v$\sim$0.11c, which corresponds to 
$R_{launch}$ $\sim$ $80R_{\rm s}$  (where $R_{\rm s} = 2GM/c^2$ is the
Schwarzschild radius), or $5\times 10^{14}$~cm for \pg.  The EPIC data show significant flux variability on timescales of 2-3 hours (figure 1 in Pounds
\et\ 2003), which would be compatible with the above scale size relating to the primary (disc/corona) X-ray emission region. 

An important prediction of the Black Hole Winds model of K03 is that the outflow momentum will be simply related to the Eddington luminosity,
viz $\mo.v = \le$ /c, an equivalence that can be tested for \pg.
Using the mass rate and velocity for the stacked data we have $\mo$.v =  $2.5\times 10^{35}$ and, for an AGN mass of $4\times 10^{7}$$\msun$ 
(Kaspi \et\ 2000), 
$\le$/c = $1.7\times 10^{35}$
- in reasonably good agreement with the continuum driving model.

The location and structure of the lower ionisation absorber, responsible for most of the continuum absorption in our model, is unclear. The
lower velocity indicated by our spectral fitting suggests a separate flow component. With an imprint only on the primary continuum, it seems that
this log$\xi$$\sim$2 matter exists close to the inner accretion disc, while the absence of substantial line emission from that matter implies a small
emission measure, possibly constrained in small, higher density clouds (e.g Risaliti \et\ 2005). This possibility could be tested for \pg\ by
more extended observations to constrain the timescale on which the primary continuum absorption is seen to change. 

In an extension of the model of K03, a possible scenario might be where an inhomogeneous flow accretes
through the inner disc to a radius R where radiation pressure causes the matter to be launched at the local escape velocity. As the outflow expands
outward the mean density will fall, as will the filling factor of the cooler, more opaque, matter. 

In conclusion, we suggest that the unusual properties of \pg\ (at least among local AGN) are due to a super-Eddington
accretion rate, a condition expected to apply to galaxies in their rapid growth phase. In that sense \pg\ might be described as a
late developer. However, we note that if the current outflow rate is maintained for $2\times 10^{7}$ years, the mechanical energy of $\sim$$10^{60}$ ergs carried into the 
host galaxy would 
exceed the binding energy for a typical galactic bulge containing $10^{11}$ stars with a velocity dispersion of 
300 km s$^{-1}$.

\section{Summary}

(1) New \xmm\ observations of \pg\ in 2007 have found the sub-relativistic outflow of highly ionised gas first seen in 2001 to be persisting.

(2) Stacking all existing EPIC data has shown a classical PCygni line profile to describe the emission and absorption components in the Fe K
band. Deconvolving those components finds a broad emission line close to the rest energy of the resonance line of FeXXV, the principal ionisation
state previously identified in the  high velocity absorption spectrum.

(3) Interpreting the emission line width in terms of velocity broadening indicates the fast outflow is occurring over a wide cone, while the 
large equivalent width of the FeXXV emission line is further direct evidence of a strong and persistent outflow of highly ionised
gas.

(4) Re-fitting the broad-band spectral model of Pounds and Reeves (2007) to the stacked data allows the time-averaged absorption and
re-emission by the high ionisation gas to be quantified, yielding a separate estimate of the covering factor of the fast outflow, of$\sim$0.3-0.6.

(5) These X-ray data provide the clearest evidence to date 
of an AGN
outflow carrying sufficient mechanical energy to disrupt starburst growth and provide the predicted coupling of the growth of a supermassive black
hole and the host galaxy.  

(6) We suggest that \pg\ is a rare example of a bright, low redshift, type 1 AGN accreting at the Eddington  limit.

\section*{ Acknowledgements }

The results reported here are based on observations obtained with \xmm, an ESA science mission with
instruments and contributions directly funded by ESA Member States and
the USA (NASA).
The authors wish to thank Valentina Braito for re-extracting the earlier \pg\ spectra and the SOC and SSC teams for organising the \xmm\
observations and initial data reduction.

\section{Appendix}

Implicit in using the stacked spectral data to determine the energy, width and strength of the emission and absorption components of the Fe K 
PCygni profile is an assumption that the derived parameters are a true representation of the time-averaged values. In particular it is important 
to confirm that the broad ionised emission line is not an artefact of the stacking procedure. 

The ratio of each pn data set individually modelled
to an absorbed power law in the energy band 3-10 keV is shown in figure 12.
Absorption and emission lines were then added to each data set, with line energy, width and flux free, and a new best fit obtained in each case. 
Table 4 below shows the 
results. Although some individual parameters are
not well constrained, and the absorption line is not formally detected in observation 2, the results are consistent with those from the
stacked data. A broad emission line is preferred in each individual data set.

\begin{figure}                                                                              
\centering                                                              
\includegraphics[width=5.9cm, angle=270]{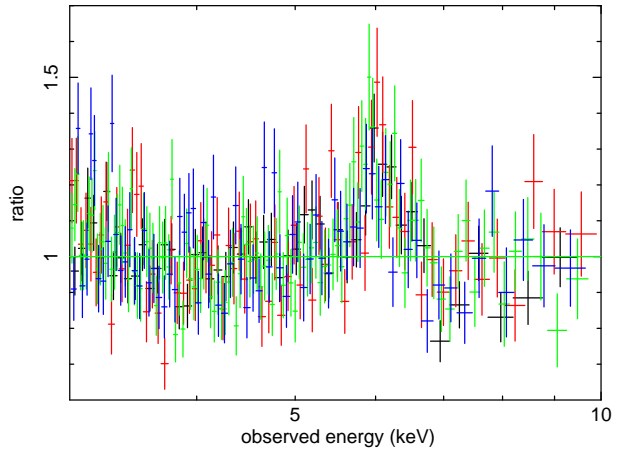}                                                               
\caption
{Fe K profiles for the individual pn data sets from the 2001 (black), 2004 (red), 2007a (green) and 2007b (blue) \xmm\ observations of \pg.
The variable data quality reflects the effective exposures, which for the pn camera were 49ks, 33ks, 44ks and 26 ks respectively}
\end{figure}

\begin{table*} 
\centering
\caption{Parameters of Gaussian absorption and emission lines fitted to the individual data sets displayed in figure 12. 
Line energies (adjusted to the AGN redshift) and 1$\sigma$ line widths are in keV and equivalent widths are in eV. The final column 
gives the improvement in $\chi^{2}$ compared with the best fit absorbed power law over the 3-9 keV band}
\begin{tabular}{@{}lccccccc@{}}
\hline
fit & absorption line energy & width & EW & emission line energy & width & EW  & $\Delta\chi^{2}$/dof \\

\hline
obs 1 & 7.62$\pm$0.05 & 0.06$\pm$0.03 & -95$\pm$30 & 6.63$\pm$0.08 & 0.33$\pm$0.10 & 135$\pm$45 &  48/6 \\
\hline
obs 2 & - & -  & - & 6.59$\pm$0.07 & 0.25$\pm$0.07 & 95$\pm$45 & 24/6 \\
\hline
obs 3 & 7.59$\pm$0.07 & 0.10$\pm$0.06 & -130$\pm$45 & 6.58$\pm$0.08 & 0.38$\pm$0.08 & 160$\pm$65 & 47/6 \\
\hline
obs 4 & 7.51$\pm$0.15 & 0.25$\pm$0.15 & -100$\pm$80 & 6.45$\pm$0.17 & 0.46$\pm$0.36 & 110$\pm$65 & 14/6 \\
\hline
\end{tabular}
\end{table*}

For the RGS data, where we found evidence for a broad emission line of OVIII, the simultaneous detection of narrow lines of OVII
provides direct support for the robustness of the stacking procedure. Again, as for the Fe XXV emission line, the large
equivalent width of the broad OVIII emission line is incompatible with a highly collimated outflow.

\end{document}